\documentclass[oneside,letterpaper,10pt,pointlessnumbers,normalheadings,halfparskip,
bibtotoc]{scrartcl}

\usepackage{xspace}
\usepackage{graphicx}
\usepackage{url}
\usepackage{color}
\usepackage{array}
\usepackage{times}
\usepackage{cite}

\newcommand{\ie}{i.\,e.,\xspace}
\newcommand{\eg}{e.\,g.,\xspace}

\newcommand{\PBS}[1]{\let\temp=\\#1\let\\=\temp}

\begin{document}

\title{\Large Tool Support for Continuous Quality Controlling}

\author{\normalsize F. Deissenboeck, E. Juergens, B. Hummel, S. Wagner, B.
Mas y Parareda, M. Pizka}

\date{\vspace{-1em}}

\maketitle

\begin{abstract}

Software systems over time suffer from a gradual quality decay and
therefore costs rise if no pro-active countermeasures are taken. Quality
controlling is the first step to avoid this cost trap. Continuous quality
assessments enable the early identification of
quality problems, when their removal is still inexpensive, and aid in
making adequate decisions as they provide an integrated view on the
current status of a software system. As a side effect, continuous and timely
feedback enables developers and maintainers to improve their skills and
thereby helps to avoid future quality defects. To make regular quality
controlling feasible, it has to
be highly automated and assessment results need to be presented in an
aggregated manner to not overwhelm users with too much data. This article
gives an overview of tools that aim at solving these issues. As an example,
we present the flexible, open source toolkit ConQAT that supports the
creation of dashboards for quality controlling and report on its
application.

\end{abstract}

\begin{tabbing}
\textbf{\textsf{Keywords:}} \= D.2.19 [Software Engineering]: Software Quality\\
\> D.2.9 [Software Engineering]: Management (Project control \& modeling)
\end{tabbing}

\paragraph{Contact Information}

\begin{tabbing}
Florian Deissenboeck, Elmar Juergens,\hspace{3cm} \= Benedikt Mas y Parareda, \\
Benjamin Hummel,  Stefan Wagner \> Markus Pizka\\
\\
Institut f\"{u}r Informatik \> itestra GmbH\\
Technische Universit\"{a}t M\"{u}nchen  \> Ludwigstra\ss e 35\\
Boltzmannstr. 3 \> 86916 Kaufering\\
85748 Garching  \> Germany \\
Germany\\
\\
\small\{deissenb, juergens, hummelb, wagnerst\}@in.tum.de \> \small \{mas, pizka\}@itestra.de\\
\end{tabbing}

\clearpage
\section{Continuous Quality Controlling Tools}

Long-lived software systems are known to undergo a gradual quality decay if
no countermeasures are
taken~\cite{1994_parnas_software_aging,2001_eick_code_decay}. Without
exception, this affects all of the quality attributes defined by ISO 9216:
reliability, functionality, efficiency, portability, usability and, above
all, maintainability. \emph{Continuous quality controlling} is used to
identify and resolve quality defects early in the development process,
while the implementation of countermeasures is still inexpensive.
Obviously, assessing the current state of a system's quality is a key
activity of quality controlling. As there is a great number of diverse
aspects of quality that need to be controlled, it must be assured that the
cost associated with these assessments does not outweigh the benefits.
Realistically, continuous quality controlling can only work in practice if
appropriate tool support is available for all automatically assessable
quality aspects. To support the selection of tools, we discuss key
requirements for quality controlling tools and present examples of tools
together with a categorization that structures the sometimes bewildering
landscape of existing tools.

\begin{figure}[h]
  \begin{center}
  \includegraphics[scale=1.1]{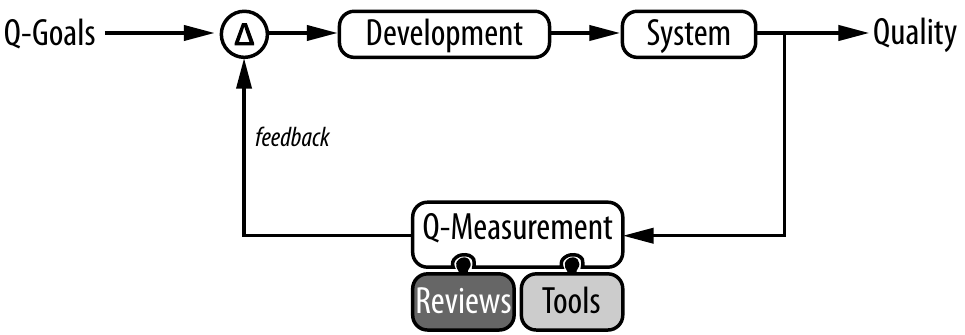}
  \caption{The Quality Controlling Loop}
  \label{fig:controlling}
  \end{center}
\end{figure}

In essence, quality controlling consists of three key elements: (1) clearly 
defined quality goals, (2) techniques, tools and processes to analyze the 
current state of quality, (3) appropriate measures to react to discovered 
quality deficits. Fig.~\ref{fig:controlling} depicts a control loop as used 
to describe dynamic systems to illustrate the process of continuous quality 
controlling. For clarity's sake, changes to the system that are not caused 
by quality defects but by changing functional requirements are not included 
in the figure. A team or person responsible for the quality of a software 
system, the quality engineer, defines the quality goals, \eg in the form of 
quality models, standards, KPIs or software metrics. The quality engineer 
uses manual techniques like reviews as well as  quality analysis tools like 
unit testing frameworks or static analyzers to \emph{measure} the quality 
of the system. He then compares the measurement results to the defined 
quality goals. Based on the outcome, he asks the \emph{developers} to 
perform the required quality improvements. These changes result in a new 
revision of the \emph{system} that again undergoes quality assessment. In 
the course of this improvement process developers are educated to increase 
quality in the long-term. If needed, the quality engineer can react to 
changing quality requirements by improving the quality goals (not 
illustrated in Fig.~\ref{fig:controlling}).

Software quality is a complex and multifaceted 
concept~\cite{1996_kitchenhamb_quality_elusive_target}. Consequently, 
definitions of quality typically differ across organizations and projects. 
However, independently of how quality is defined in a particular context, 
quality measurement is required to determine if a system meets the defined 
quality goals. Due to the number of relevant quality criteria and the size 
of the analyzed systems, the measurement in itself is a complex and 
challenging task. Hence, this article does not discuss specific quality 
criteria but focuses on the tools that are needed to monitor quality and 
thereby make continuous quality controlling effective and efficient.

There is a plethora of quality analysis tools offered by commercial vendors 
as well as academia that credit themselves with the ability to accomplish 
this. The range goes from unit testing frameworks over metric tools, 
violation checkers, architecture assessors to dashboards, software cockpits 
and application intelligence platforms. However, it is challenging to 
decide on the \emph{right} tool for quality controlling as the requirements 
for such a tool are often unclear. To support decision making, we propose 
to use as a basis the following list of requirements, that, according to 
our experience, are substantial for putting continuous quality controlling 
into practice.

\textbf{Aggregation \& Visualization.} Automated quality analysis of large
software systems generates an enormous amount of analysis data. To not
overwhelm users, analysis results need to be aggregated to a comprehensible
level and presented in an appropriate manner. Aggregations can be achieved
through categorization of metrics based on thresholds, or through basic
aggregation operators like maximum, average, median etc. As results like
``the average cyclomatic complexity of the system is 15'' are of limited
practical use, the latter aggregation approach requires the tool to be
aware of the analyzed system's structure to produce meaningful results.
Even highly aggregated analysis results need to be conveyed to the user in
an effective manner. Hence, the tool must provide powerful visualization
mechanisms beyond tables and charts, \eg graphs for structural information
or tree maps to visualize distribution of anomalies within a system.

\textbf{Dedicated Views.} The assessment results must be accessible to all
project participants, \eg on a website or via a specific client. However,
due to differing interests, it must be possible to provide a customized
view for each stakeholder. Project managers, for example, are mainly
interested in a high-level overview that enables them to spot problems
without going into details. Developers, on the other hand, require views of
finer granularity that allow them to inspect analysis
results for the artifacts they have been working on.

\textbf{Trend Analysis.} Many quality defects are hard to identify by
investigating a single snapshot of a system but can be discovered by tracking
changes over time. Hence, the tool must be capable of storing and presenting
historical data to foster the identification of trends. Moreover, various
metrics used today are hard to interpret on an absolute scale but are
well-suited if relative measures are used. It is, for example, not entirely
clear what the quality implications of a code cloning ratio of 16\% are, whereas most quality
engineers would agree that it is important to ensure that the cloning ratio
does not increase over time. (The cloning ratio measures how much of the source
code has been copied at least once. It thus provides an estimate of how likely
changes will have to be performed in multiple places due to code duplication.)

\textbf{Customizability.} Quality requirements are highly project-specific 
as the analyzed systems, the applied tools and processes, the involved 
technologies and the acting people differ. Even more so, these requirements 
are not constant but evolve over the course of a project. Hence, quality 
analysis tools must be highly customizable to support a project-specific 
tailoring of the analyses carried out and the way they are presented. 
Besides this, false positives generated by analysis tools are known to be a 
severe obstacle to the acceptance of continuous quality controlling as they 
cause frustration among users. Here, customization is required to configure 
analysis tools to reduce the number false 
positives~\cite{2008_wagnerst_bug_patterns}. For example, defects that 
impact code readability are not interesting for developers if they occur in 
generated code that is never actually read. In such cases, analyses must be 
tailored to be aware of generated code.

\textbf{Diversity.} The factors influencing product quality are diverse.
Therefore a quality controlling tool may not be limited to a certain type
of factors or artifacts it analyses. It must not only be able to analyze
source code but should provide measures for other artifacts like
documentation, models, build scripts or information stored in a change
management systems. As quality attributes can be discussed on many
different levels, the tool should make no restrictions on the level of
detail, the level of granularity, nor the type of analysis. It must, for
example, be possible to analyze a source code artifact on representation
levels as different as character stream, token stream, syntax tree or call
graph.

\textbf{Extensibility.} As no tool can innately support the whole spectrum
of all possible artifacts it must provide an extension mechanism that
allows users to add further analysis or assessment modules as needed.
Examples of such artifacts are models used in model-driven development or
programs written in newly created domain-specific languages.

\textbf{Non-interactive operation.} Tool supported assessments need to be
carried out regularly (\eg hourly or daily) to provide timely results. To
achieve this in a cost-efficient manner, analysis tools need to be able to
work in a completely automated, non-interactive way.

\textbf{Performance.} As quality controlling is particularly relevant for large
scale systems that prevent comprehensive manual assessments, a quality
controlling tool must be able to cope with the analyzed system's size in
acceptable time.

\section{Tool Landscape}
Since the tasks and activities related to quality analysis and controlling are
manifold and diverse, the landscape of existing tools is correspondingly large.
In order to simplify orientation and selection of suitable tools for a specific
purpose, we propose a classification of quality analysis tools according to
usage-related dimensions. More specifically, we outline characteristics and
representatives of four tool categories, namely \emph{Sensors}, \emph{System
Analysis Workbenches}, \emph{Project Intelligence Platforms} and
\emph{Dashboard Toolkits}. We are well aware that the boundaries of these
categories are blurred. If a tool can be argued to belong to more than one
category, we tried to classify it according to its primary use case.
Furthermore, due to the large number of existing tools and their pace of
evolution, the list of representatives for each categorization is inherently
incomplete, which we attempted to alleviate by choosing representative examples
for each category. Nonetheless, we are convinced that such a categorization is
fundamental as a guide in the diversity of the tool landscape and to
distinguish between different use cases. Table \ref{categorization} gives an
overview of the categories and corresponding tools that are further detailed in
the following paragraphs.

\begin{table}[htbp]
{\footnotesize\sffamily
\begin{tabular}{>{\PBS\raggedright}p{0.11\linewidth}|>{\PBS\raggedright}p{0.2\linewidth}|>{\PBS\raggedright}p{0.175\linewidth}|>{\PBS\raggedright}p{0.18\linewidth}|>{\PBS\raggedright}p{0.2\linewidth}}
 & \textbf{Sensors} & \textbf{System Analysis Workbenches} & \textbf{Project Intelligence Platforms} & \textbf{Dashboard Toolkits}
 \\\hline
 scope & quality analysis & quality analysis & project controlling \& quality analysis & quality analysis \& project controlling \\\hline
 interaction paradigm & autonomous & interactive & autonomous & autonomous\\\hline
 usage scenario & nightly-build, IDE integration & demand-driven & demand-driven \& nightly-build & nightly-build \\\hline
 analysis object & development artifacts & code, architecture & metrics & project \& process artifacts\\\hline
 analysis question & hard-wired & queries on system snapshot & queries on metric data & configuration of analysis topology\\\hline
 result represent. & lists & artifact-specific visualizations & lists, charts & list, charts \& artifact-spec. visual.\\\hline
 examples & JDepend, PMD, FxCop, NDepend, PC-Lint, Klocwork, JUnit & Sotograph,  iPlasma & Hackystat, Team  Foundation Server & ConQAT,  XRadar, QALab, Sonar
\end{tabular}}
\caption{Use case driven tool categorization}\label{categorization}
\end{table}

\textbf{Sensors} comprise verification and testing tools, anomaly detectors and
metric calculators that perform fully automated analyses of development
artifacts w.r.t.\ specific quality criteria. Due to their autonomous
interaction paradigm that requires no user input, a common usage scenario is
their application during automated nightly builds or as compile-time checkers
in modern IDEs. The type of analysis question they answer is typically
hard-wired. Analysis results are presented as tables or lists or as markers
within an IDE. Examples include \emph{JDepend} \cite{jdepend} that computes
dependencies between components and warns if certain dependency rules are
violated and \emph{PMD} \cite{pmd} that performs guideline checks and bug
pattern search for Java programs. \emph{NDepend} \cite{ndepend} and
\emph{FxCop} \cite{fxcop} are representatives of comparable tools for the .NET
platform and \emph{PC-lint} \cite{pclint} or \emph{Klocwork} \cite{klocwork}
perform, amongst others, guideline checks and inspections of security
vulnerabilities for C/C++. Moreover, \emph{xUnit} frameworks \cite{junit}
automate unit tests for numerable programming languages.

\textbf{System analysis workbenches} support experts in the analysis of various
development artifacts, including source code or architecture specifications, in
order to answer analysis questions about specific quality aspects of a system,
such as its architecture conformance or component structure. In contrast to
sensors, they are interactive tools that are used on demand, during
system inspection or review. They support interactive analysis by offering
flexible system query languages and present results using specialized
artifact- and task-specific visualizations, including graphs, charts and tree
maps. \emph{Sotoarc/Sotograph} \cite{sotoarc} and \emph{iPlasma} \cite{iplasma} are
commercial respectively open source products for comprehension and reverse
engineering of software systems. They provide analysis middle-ware in the form
of a repository with a fixed metamodel into which systems under inspection are
loaded for convenient access. Sotoarc supports modeling of a system's intended
architecture and evaluation of the architecture conformance of its
implementation. Furthermore, it can simulate restructurings to evaluate effects
of architecture modifications. iPlasma offers, beside architecture analyses, a
suite of object-oriented metrics and duplication detection. Furthermore, it
provides a language to specify static analyses and a visualization framework
and can thus be used as a basis for the development of further interactive
analyses.

\textbf{Project intelligence platforms} collect and store product and process
related metrics of multiple sources to perform trend or comparative analyses.
They are deeply integrated into a software development environment
and collect metric data as it originates during development. Flexible query
mechanisms often support generation of reports that show charts
depicting the evolution of selected metric values over time. If these reports
are generated in a frequent manner, they can serve as a project dashboard. Even
though query creation has an interactive nature, project intelligence platforms
operate autonomously and collect metric data without developer
interaction. Project intelligence platforms are usually limited to metric
values and allow ad-hoc queries on data from the project's past.
\emph{Hackystat} \cite{hackystat} is an open source framework for collection
and analysis of software development process and product data. It originated
from the work of the Collaborative Software Development Laboratory at the
University of Hawaii to support software project telemetry
\cite{2005_Johnson_hackystat}. Hackystat offers sensors that gather
data during software development and transmit it to a central
server for analysis, aggregation and visualization. Via a custom query
language, reports can be specified to visualize, correlate or compare measured
data. This way, hypotheses about the development process can be tested and
impact of process changes on project performance can be evaluated. Although
Hackystat offers several sensors that interface with static analysis tools that
perform product quality analyses, its emphasis is on process
measurement. \emph{Microsoft Team Foundation Server} \cite{tfs} is a commercial
software product that aims to support collaborative software engineering.
Besides source control and issue tracking functionality, it provides data
collection and reporting services. Collected source control, issue tracking,
build results, static analysis and test execution data is stored in a relational
database system from which a reporting engine generates reports that monitor
process metrics and visualize trends. The emphasis of the data collection and
reporting services is on process related data collection and reporting.

\textbf{Dashboard toolkits} provide libraries of building blocks from which
custom-made analysis dashboards, that collect, relate, aggregate and visualize
sensor data, can be assembled by configuration. Building block libraries offer
sensors for analysis of both product (\eg code, architecture) and process (\eg
source control or issue management information) related artifacts.
Additionally, blocks for presentation allow analysis results to be represented
in a variety of formats, including general purpose lists or tables and
specialized visualizations such as trees, graphs, charts or tree maps. In
contrast to system analysis workbenches that are geared towards an interactive,
on-demand explorative analysis of a system snapshot, dashboard toolkits are
used for continuous quality analysis and monitoring of a project-specific set
of questions. Their scope thus comprises both quality analysis and project
controlling. In contrast to project intelligence platforms, that focus on
operations on numerical metric values, dashboard toolkits can access sensor
information on the level of development artifacts. They can thus exert greater
control over sensor operations, which facilitates customization of analyses to
project specific settings. Dashboard toolkits can be differentiated by the
degree of customizability they provide. Both \emph{QALab} \cite{qlab} and
\emph{Sonar} \cite{sonar} offer pre-configured dashboards that present output
from various sensors  but offer very limited customization capabilities. QALab
creates trend analysis charts displaying the evolution of the number of
anomalies of a project. Sonar provides additional visualizations displaying
aggregated single-project or cross-project quality information. Customization
capabilities of both tools are limited to the choice of applied sensors.
\emph{XRadar} \cite{xradar} is an open source code report tool for Java-based
systems, which integrates XML reports calculated by different sensors via XSLT
transformations. The results can be aggregated along the package hierarchy and
also stored in the file system for plotting trend graphs of various metrics.
Due to the expressiveness of XSLT transformations, aggregation and
visualization of the imported analysis results can be configured more flexibly,
than with QALab or Sonar. \emph{ConQAT} \cite{conqat} is an open source dashboard 
toolkit that offers a rich library of analyses, filters, aggregations and 
visualisazions for systems written in various languages. Due to its pipes-and-filters 
style configuration mechanism, it offers high extensibility and flexibility 
for project-specific tailoring. ConQAT is described in more detail in the sidebar.

\begin{table}
{\small
  \begin{tabular}{|p{\textwidth}|}
    \hline
    \multicolumn{1}{|c|}{\large Sidebar:
      ConQAT: The Continuous Quality Assessment Toolkit} \\
    \hline

The \emph{Continuous Quality Assessment Toolkit} ConQAT
\cite{2005_deissenboeckf_conqat} was developed over the last
three years at the Technische Universit\"at M\"unchen within the scope
of multiple industrial projects and released as open source in 2007.
The rationale behind developing our own dashboard toolkit was that no
available tool covers all of the requirements we consider to be
essential and we wanted to have a flexible and extensible platform
for our projects and experiments.

ConQAT employs a \emph{pipes-and-filters}-architecture to achieve the
requirements of customizability and extensibility.  Configurable units
dedicated to specific analysis tasks (such as preprocessing,
assessment, aggregation, visualization etc.) are connected in a
dataflow diagram for achieving the desired results. An example of this
is shown in the screen-shot of the accompanying graphical editor. A
complex control center monitoring multiple projects can easily consist
of hundreds of those units, which we call \emph{processors}. To
simplify the configuration process, reoccurring patterns of processors
can be encapsulated as a \emph{block} which then can be instantiated
multiple times.

% Note: 253 procs+blocks, 44 blocks
The current version of ConQAT provides more than 200
processors and 40 blocks, and can analyze such diverse artifacts as
source code (Java, C\#, C++, PL/1, Cobol and VB), configuration management
systems or Matlab Simulink models. Analysis capabilities range from the
usual basic metrics to complex assessments, such as detecting
violations of the software architecture. The most important aspect
however is the aggregation and presentation of the collected data,
which has to make an overwhelming amount of information accessible to
the user. Thus ConQAT includes configurable processors for aggregation
and visualization using diagrams, tree maps and layouted graphs. To
allow trend analysis of key measurements, all analysis results can be
persisted in a database.

Wherever possible, approved libraries and sensors (such as PMD or
Klocwork) have been integrated or interfaced, while for advanced tasks
for which the currently available tools did not seem mature enough
(such as quality analysis of models) or which are still subject to
active research (\eg model clone detection \cite{icse08}) individual
solutions are implemented. If a desired assessment value can not be
calculated using the facilities already provided with ConQAT, custom
processors can be written in Java. Most of the time these are
short classes, as pre- and post-processing tasks and
visualization is already handled by existing processors.

As we consider the customization of a dashboard to the current project
context to be one of the key requirements, a graphical editor (shown
in the screen-shot) for creating and modifying analysis configurations
is available. This way navigating and validating these configurations
is greatly simplified, making the creation of control centers more
intuitive and efficient.

\includegraphics[width=9cm]{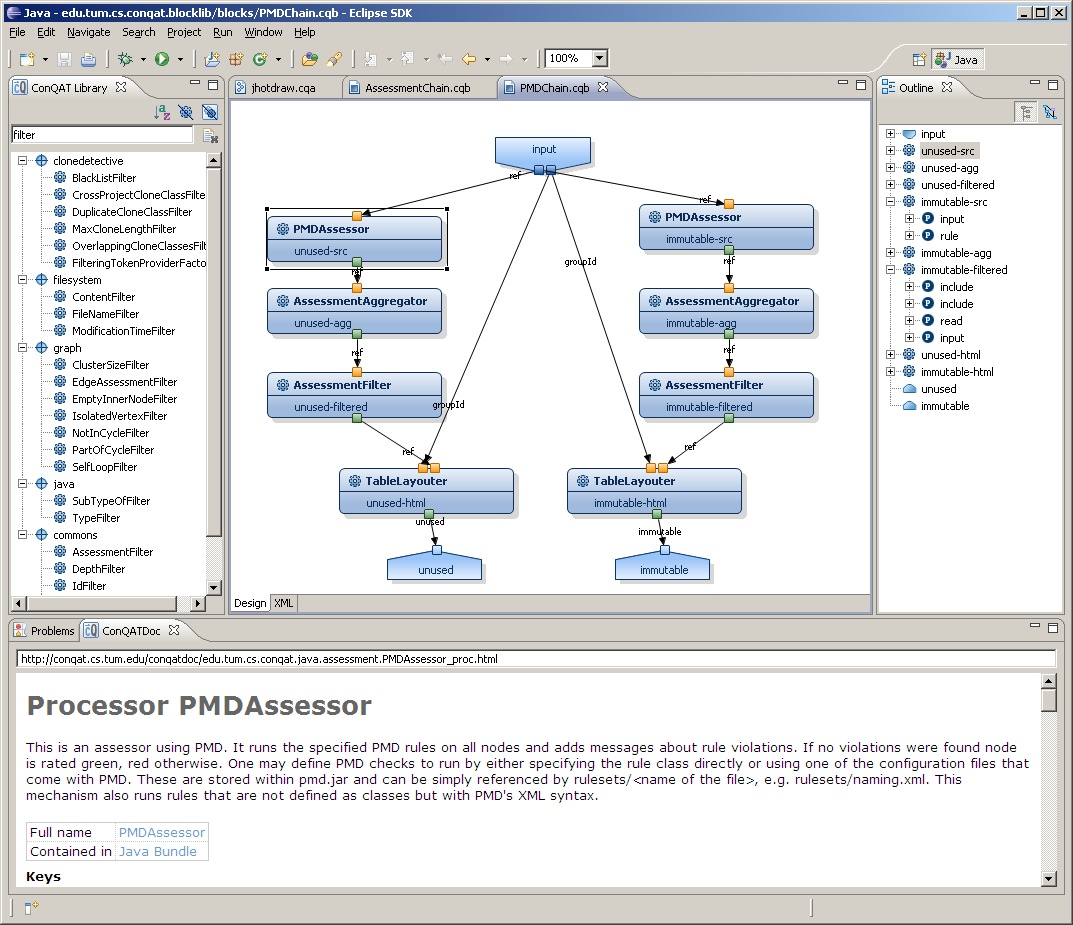}

    \\
    \hline
  \end{tabular}}
\end{table}

\section{Experiences}

We have worked in various academic as well as industrial projects on 
different aspects of quality controlling. Tool support has always been an 
important topic in these projects. Hence, the presented requirements and 
categorization stem largely from these experiences although they have, in 
parts, also been identified by others \cite{2007_Johnson_hackystat}. We use 
our own dashboard toolkit ConQAT as main example to report in 
more detail on our practices with tool support for continuous quality 
controlling.

The re-insurance company \emph{Munich Re Group} runs a portfolio of
diverse, large and long-lived business systems based on the .NET platform. For
these, a
ConQAT-based quality dashboard has been installed which monitors key
maintainability measures such as redundancy and compliance to the
system's architecture and is also used to support manual reviews.
Currently it is employed for maintenance and further development of a
business system of about 400 kLoC. In the controlling process the
identified clones and architecture violations have been reduced based
on the tool results. As we observed that the output of
individual sensors is often too detailed to identify relevant information, a
substantial amount of work has been put into the
configuration of suitable aggregators and filters. This has been
decisive in order to make only the most relevant problems visible. The
lesson learned was that tailorability is a crucial feature of a dashboard.

A somewhat different analysis scenario was encountered by MAN, a truck
manufacturer, where continuous quality controlling was to be applied
to a model-based development process of embedded systems.  Major parts
of the power train controller are modeled using Matlab
Simulink/Tar\-get\-Link.
The entire implementation is generated from these
models.
The main object of analysis was not source code but
models and the quality analysis tool had to be adapted to that new
artifact type.
The employed analyses included the collection of various size measures
such as number of blocks and states in diagrams or identification of
the usage of blocks not supported by the code generator.
It showed that easy extensibility is important in order
to cope with the variety of artifacts encountered.
Additionally, as we could
reuse existing parts for
graph analysis and layout, we see flexible libraries in dashboards
as a crucial support for efficient analysis development.

The IT-consulting company \emph{itestra} performs quality analyses of very 
large mainframe applications with the help of ConQAT. These software 
assessments influence decisions on future development, maintenance and 
operation of software systems.  The source code-related measures are 
collected, aggregated and inserted into a history database. So far, itestra 
analyzed systems containing more than 20 MLoC. In each assessment, more 
than 60 measures are collected, including size statistics (source lines of 
code, number of procedures, average procedure length), structural data 
(condition ratio, maximal loop nesting), and redundancy measures (code 
cloning). The size of these business systems as well as the large number of 
measures required a scalable tool that allows efficient aggregation of 
these numbers. Additionally, the systems under evaluation are mainly 
written in COBOL or PL/1, languages for which only scarce tool support 
exists. Hence, high extendability as well as aggregations and 
visualizations were needed to provide a viable foundation for the 
development of a suitable analysis tool.  Only the core parts for each 
analysis had to be developed, while all of the pre-processing, aggregation, 
and visualization support of ConQAT integrated seamlessly.

At Technische Universit{\"a}t M{\"u}nchen we apply continuous quality
controlling in programming lab courses not only to monitor product
quality but also to give immediate feedback to students. In this
context, non-interactive operation is crucial to be able to present
students with frequent feedback without straining supervisors'
resources. Moreover, dedicated views are required to provide
supervisors with an overview of the whole project while presenting
students with feedback on their own work only. Besides this, ConQAT is
applied on itself in order to control its own quality.

It is our collective experience that,
while each of the tools outlined in the tool classification provides
individual value, none of them fully satisfies the above mentioned requirements
for continuous quality controlling. Individual sensors are limited to
the analysis of specific quality criteria and hence cannot provide
comprehensive quality
assessments. The interactive usage paradigm of system analysis workbenches
makes them unsuited for continuous application during a nightly build. 
Project intelligence platforms work on data collected by external sensors. 
Since customization to project specific settings frequently requires 
adaptation of sensor functionality, their customizability is limited. 
Finally, the outlined dashboard toolkits are not customizable enough to 
satisfy the extensibility and diversity requirements. 
Our highly configurable dashboard toolkit ConQAT is a first 
step to alleviate these shortcomings.

\section{Conclusion}
As stated in the beginning of this article, successful quality controlling
starts with knowing what to measure; \ie the definition of relevant
quality goals. Though this sounds evident, it has to be stressed because
software quality as a technical discipline is all but
mature~\cite{1996_kitchenhamb_quality_elusive_target}. In contrast to
mature industries and despite of the enormous spendings for software there
are still neither legally binding nor broadly accepted standards for
software quality. The question what is good and bad in software is yet
mostly unanswered and forces organizations to find individual innovative
quality rules and guidelines.

In addition to individual quality models, tool vendors also have to deal
with the tough technical difficulties of (static) software analysis.
Writing parsers for programming languages such as C/C++ or even PL/I is
already very challenging. However, building effective analyzers requires
even stronger efforts, especially for the analysis of the more relevant
semantical properties of a software system based on sound heuristics and
estimates. Due to these difficulties, some tools unfortunately measure what
is easy to measure but not what was important to know.

Due to the combination of a) the individual character of quality rules and
b) the difficulties to implement effective analyzes it is currently very
unlikely that a single tool of a certain vendor can adequately support all
quality controlling needs of an organization.

In many cases, more than one tool will be needed and the user might want to
add his or her own quality rules. The tool categories and requirements
outlined in this article can serve as a starting point for selecting
appropriate tools. A flexible and extensible dashboard toolkit like ConQAT 
seems to be an attractive technical foundation for quality controlling. 
It allows to quickly gain benefit from quality controlling with low initial 
overhead by observing few important properties and communicating them 
frequently to managers, project leaders and developers. Over time it can 
be evolved incrementally with the sharpening perception of software quality 
within the organization.

However, it must be pointed out that, albeit crucial, tool-support is only 
one aspect of quality controlling. A prerequisite for successful quality 
controlling is an organization's readiness to assess and improve quality 
although many quality deficits have long-term and therefore less obvious 
consequences. From our experience, this readiness is put to test as soon as 
it comes to manual quality assessments. Studies again and again show that 
manual review activities pay off. Still, organizations often try to 
avoid them as, from a short-term perspective, they are perceived as being 
too costly. As many essential quality issues, such as the usage of 
appropriate data structures and meaningful documentation, are semantic in 
nature and can inherently not be analyzed automatically, we consider this a 
precarious situation.

% -------------------------------------------------------------------------
\bibliographystyle{unsrt}

\newpage
\section*{Biographies}

\paragraph{Florian Deissenboeck}
is a research assistant at the Software \& Systems Engineering group of 
Prof.~M.~Broy at the Technische Universit{\"a}t M{\"u}nchen. Currently he works 
on his PhD thesis about software quality controlling. His academic 
interests lie in software maintenance, software product quality and program 
comprehension. He studied computer science at the Technische Universit{\"a}t 
M{\"u}nchen and the Asian Institute of Technology, Bangkok.

\paragraph{Elmar Juergens}
works as a research assistant and PhD student at the Software \& Systems 
Engineering group of Prof.~M.~Broy at the Technische Universit{\"a}t M{\"u}nchen. 
His academic interests include software maintenance, clone detection and 
usage analysis. He studied computer science at the Technische Universit{\"a}t 
M{\"u}nchen and the Universidad Carlos III in Madrid, Spain.

\paragraph{Benjamin Hummel} is a research assistant and PhD student at
the Software \& Systems Engineering group of Prof.~M.~Broy at the
Technische Universit\"at M\"unchen. His research interests are
modeling and verification of discrete and hybrid systems, and software
quality and maintenance. He received a diploma in computer science
from the Technische Universit{\"a}t M{\"u}nchen and is a member of the
ACM and the Gesellschaft f\"ur Informatik (GI).

\paragraph{Stefan Wagner} works as a post-doctoral researcher at the 
Software \& Systems Engineering group of Prof.~M.~Broy at the Technische 
Universit{\"a}t M{\"u}nchen. His main interests include quality modeling and
analysis, especially the connection to economics. He studied 
computer science in Augsburg and Edinburgh and
holds a doctoral degree in computer science from the 
Technische Universit\"at M\"unchen. He is a member of ACM SIGSOFT, the 
IEEE Computer Society, and the Gesellschaft f\"ur Informatik (GI).

\paragraph{Benedikt Mas y Parareda} is an IT consultant researcher working for
itestra GmbH, a spin-off of the Technische Universit{\"a}t M{\"u}nchen.
He specializes in software assessment and quality evaluation. Mr.~Mas y Parareda
studied Computer Science at the Technische Universit{\"a}t M{\"u}nchen and 
the United
Nations University in Macau.

\paragraph{Markus Pizka} is co-founder and managing director of the German
IT consultancy itestra GmbH. He received a Dr.~degree from the 
Technische Universit{\"a}t M{\"u}nchen for his work on Distributed
Operating Systems. He later shared his experiences in compilers and reverse
engineering with Microsoft Research, Cambridge and went on as a project
leader for a large German software company. He then worked as an assistant
professor at the Technische Universit{\"a}t M{\"u}nchen where he established a
competence center for software maintenance.

\end{document}